\documentclass[reprint, prmaterials, amsfonts, amssymb, amsmath, aps, floatfix, article]{revtex4-2}
\usepackage[pdftex, pdftitle={Article}, pdfauthor={Author}]{hyperref}
\usepackage{float}
\usepackage{graphicx}
\usepackage{hyperref}
\hypersetup{
    colorlinks=true,
    linkcolor=blue,
    filecolor=blue,
    citecolor=blue,
    urlcolor=blue,
    pdftitle={Overleaf Example},
    pdfpagemode=FullScreen,
    }
\usepackage{longtable}

\usepackage{xpunctuate}
\DeclareRobustCommand\etal{\xperiodafter{\emph{et al}}}

\begin{document}

\title{Minority-spin conducting states in Fe substituted pyrite CoS$_2$}
\author{Anustup Mukherjee and Alaska Subedi}
    \affiliation{CPHT, CNRS, \'Ecole Polytechnique, Institut Polytechnique de Paris, 91128 Palaiseau, France}

\date{\today} 

\begin{abstract}
There has been a longstanding debate whether the pyrite CoS$_2$ or its alloys 
with FeS$_2$ are half metallic.  We argue using first principles calculations 
that there is a finite occupation of minority-spin states at the Fermi level 
throughout the 
series Co$_{1-x}$Fe$_x$S$_2$.  Although the exchange-correlation functional 
influences the specifics of the electronic structure, we observe a similar 
trend with increasing Fe concentration in both LDA and GGA calculations.  
Specifically, even as band filling is decreased through
% electrons are removed through 
Fe substitution, the 
lowest-lying conduction band in the minority-spin channel broadens such that 
these states keep getting lowered relative to the Fermi level, which is 
contrary to the expectations from a rigid band picture.  Furthermore, the 
exchange splitting decreases as more Co atoms are replaced by Fe, and this 
again brings the minority-spin states closer to the Fermi level. These two 
mechanisms, in conjunction with the experimental observation that
minority-spin bands cross the Fermi level in stoichiometric CoS$_2$, indicate 
that minority-spin charge carriers will always be present in Co$_{1-x}$Fe$_x$S$_2$.

\end{abstract}


\maketitle

\section{Introduction}

The pyrite CoS$_2$ and related alloys Co$_{1-x}$Fe$_x$S$_2$ have been extensively investigated
due to their possible half-metallic behavior \cite{Leighton2007}.
This was motivated by the observation of Jarrett \etal that Co$_{1-x}$Fe$_x$S$_2$ is 
ferromagnetic with a saturation magnetization $M_s$ $\sim 1 \mu_B$/Co throughout the series 
$0 \leq x \leq 0.9$ ~\cite{Jarrett1968}.  However, there is a strong debate whether the
electronic states at the Fermi level exhibit full spin polarization in these compounds. 
First principles calculations within the local density approximation (LDA) by Zhao \etal 
find a small minority-spin electron pocket in CoS$_2$ \cite{Zhao1993}, making this 
material almost, but not quite, half metallic.  On the other hand, calculations within the 
generalized gradient approximation (GGA) by Shishidou \etal show the states at the Fermi
level are fully spin-polarized in CoS$_2$ \cite{Shishidou2001}.  Reflectivity measurements 
and de Haas-van Alphen experiments 
suggest that the minority-spin pocket is unoccupied \cite{Yamamoto1999,Teruya2016}. In
contrast, point-contact Andreev reflection (PCAR) experiments find a spin polarization $P$ of only 
64\% \cite{Wang2004, Wang2006b}.  Polarized neutron diffraction experiments show that occupancy of 
the minority-spin $e_g$ states is required to describe the magnetization distribution in the
ferromagnetic phase \cite{Brown2005}.  Photoemission spectra also suggest that some 
minority-spin $e_g$ states lie below the Fermi level \cite{Takahashi2001}, and this
has been confirmed by subsequent angle resolved photoemission spectroscopy (ARPES) 
experiments \cite{Wu2007, Schroter2020, Fujiwara2022}.

Although pure CoS$_2$ may not be half metallic, Mazin has argued using density functional 
calculations that Co$_{1-x}$Fe$_x$S$_2$ should exhibit half metallicity that is robust 
to crystallographic disorder for concentrations $0.10 \lesssim x  \lesssim 0.75$ \cite{Mazin2000}. 
This is because the exchange splitting of the Co $e_g$ states is large, and the gain 
in Hund energy should be greater than the cost of kinetic energy when only the majority-spin 
Co $e_g$ states are occupied as electrons are removed through
Fe substitution.  Indeed, experiments find that the total magnetization increases 
from a value of $M_s \approx 0.85 \mu_B$/Co for $x = 0$ to the integral value of 
1 $\mu_B$/Co for $x \approx 0.1$ and maintains this value until $x \approx 0.5$--$0.6$, 
before decreasing again as Fe concentration is increased \cite{Jarrett1968, Ramesha2004, Wang2005, Wang2006}.
However, PCAR experiments by Cheng \etal on samples with $0.35 \leq  x  \leq 0.9$  
show $P$ of only 47--61\%  even when $M_s = 1 \mu_B$/Co, and those
by Wang \etal find a maximum $P$ of 85\% for $x = 0.15$ \cite{Wang2005, Wang2006}.
The lack of full spin polarization is inconsistent with the results of most density 
functional calculations that consider the same levels of Fe substitution 
\cite{Mazin2000, Ramesha2004, Wang2005, Umemoto2006}, although 
one study using the supercell approach does find that minority-spin states are occupied 
within the GGA \cite{Feng2018}.
On a related note, we point out that a recent first prinicples study finds that 
hole-doped FeS$_2$ may exhibit full spin-polarization at the Fermi level \cite{Lei2021}.

\begin{table*}[!htbp]
% \centering
\caption{Relaxed metal-sulphur ($M$-S) and sulphur-sulphur (S-S) bond distances 
obtained using LDA.  We used experimental lattice constants $a$ for the end member
compounds CoS$_2$ and FeS$_2$ and interpolated ones using Vegard's law for the
intermediate compositions.  Experimental S-S distances for corresponding 
compositions are given in the final column.}
\label{Lattice parameters table}
\begin{ruledtabular}
\begin{tabular}{l c  c c c c}
% \hline
    Compound & \textit{a} (\AA) & Co-S (\AA) & Fe-S (\AA) & S-S (\AA) & S-S (\AA) exp\\ \hline
    CoS$_2$ & 5.539  \cite{Hebert2013}\phantom{, 00} & 2.316 & $-$ & 2.191 & 2.12~\cite{Nowack1991} \\
    Co$_{0.75}$Fe$_{0.25}$S$_2$ & 5.503 \cite{Cheng2003}\phantom{, 00} & 2.315, 2.302, 2.309  & 2.269 & 2.168, 2.215 & 2.115~\cite{Ramesha2004}\\
    Co$_{0.5}$Fe$_{0.5}$S$_2$ & 5.475 \phantom{[00, 00]} & 2.300, 2.308, 2.293 & 2.257, 2.269, 2.266 & 2.250, 2.228 & 2.1~\cite{Ramesha2004} \\
    Co$_{0.25}$Fe$_{0.75}$S$_2$ & 5.443 \phantom{[00, 00]} & 2.293 & 2.254, 2.265, 2.252 & 2.253, 2.259 & $-$\\
    FeS$_2$ & 5.407  \cite{Umemoto2006, Crystal} & $-$ & 2.254 & 2.208 & 2.15~\cite{Finklea1976,Nowack1991} \\
 % \hline
\end{tabular} 
\end{ruledtabular}
\end{table*}

In this paper, we reexamine the robustness of half metallicity in 
Co$_{1-x}$Fe$_x$S$_2$ using density functional calculations for 
$x = 0, 0.25, 0.5, 0.75,$ and 1.  We find that 
the details of the electronic structure and magnetism depend on whether 
one uses the GGA or LDA.  However, both approaches find that 
Co$_{1-x}$Fe$_x$S$_2$ is not a half-metal for $x > 0 $, and they 
display a consistent trend of qualitative changes in the electronic 
structure.  As electrons are removed via Fe substitution, the band width 
of the lowest-lying minority-spin conduction band increases. The dispersive 
part of the band has predominantly S $p$ antibonding character, suggesting 
that changes
in screening modifies the overlap of antibonding orbtials situated at 
neighboring S$_2^{2-}$ dimers. When Fe concentration is further increased, 
the exchange splitting of the conduction manifold comprised of Co $e_g$
and S $p$ antibonding states decreases.  Both these effects work to ensure
that minority-spin states are always occupied for Fe concentrations 
$x < 1$.  Considering the experimental fact that the lowest-lying
conduction band in CoS$_2$ is occupied in the minority-spin channel, our work implies
that Co$_{1-x}$Fe$_x$S$_2$ may not be half-metallic for any value of $x$.

\section{Computational Details}

We performed first principles calculations within the framework of density 
functional theory~\cite{Kohn1964, Kohn1965} using the Vienna \textit{Ab initio} 
Simulation package \cite{Kresse1993, Kresse1996a, Kresse1996b}. We used both 
the LDA and GGA scheme of Perdew, Burke, 
and Ernzerhof \cite{Perdew1996} to approximate the exchange-correlation 
functional.  A converged basis-set energy cut-off of at least 420 eV and 
$\Gamma$-centered $k$-point mesh of at least $8 \times 8 \times 8$ were 
used in the calculations.  The energy convergence criterion for the 
self-consistent cycles was set to $10^{-8}$ eV.  The valence electronic 
configurations of the pseudopotentials were $3d^{8}4s^{1}$ (Co), $3d^{7}4s^{1}$ (Fe) 
and $3s^{2}3p^{4}$ (S). 

\section{Structural Details}

\begin{figure}[!htbp]
    \includegraphics[width=\columnwidth]{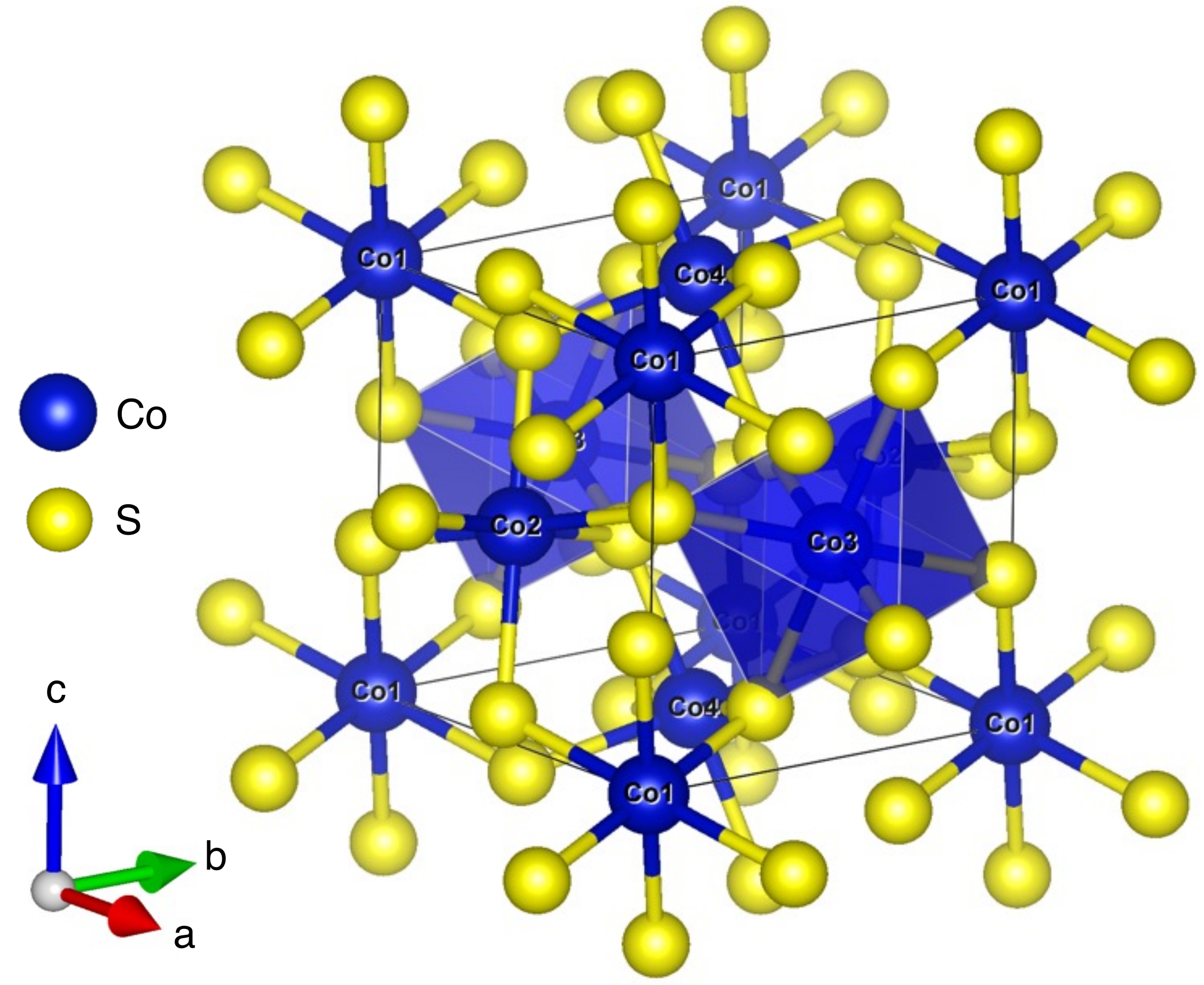}
    \caption{Structure of the unit cell of pyrite CoS$_2$ with space group $Pa\overline{3}$. 
    Nearest-neighbor Co-S distances are depicted using solid lines to illustrate the octahedral
    coordination of Co atoms. There are four formula units in the unit cell.  The sites
    labeled Co1, Co2, Co3, and Co4 are sequentially substituted by Fe to obtain 
    Co$_{1-x}$Fe$_x$S$_2$ for $x$ = 0.25, 0.5, 0.75, and 1. }
    \label{fig:CoS2_structure}
\end{figure}

Both CoS$_2$ and FeS$_2$ occur in the cubic pyrite structure with space group 
$Pa\overline{3}$, which is shown in Fig.~\ref{fig:CoS2_structure}. 
The unit cell comprises of four Co/Fe atoms at the Wyckoff position $4a$ (0, 0, 0)
 % , (1/2, 1/2, 0), (1/2, 1/2, 0) and (0, 1/2, 1/2), 
and eight S atoms at  $8c$ $(u, u, u)$, and the respective symmetry-equivalent sites.
 % ($\pm u$, $\pm u$, $\pm u$), ($\pm u$+1/2, $\pm u$, $\mp u+1/2$),  
 % ($\pm u$, $\mp u+1/2$, $\pm u+1/2$) and ($\mp u+1/2$, $\pm u+1/2$, $\pm u$).
The Co/Fe atoms are situated inside corner-shared octahedra formed by six 
dimerized S atoms.  Each S atom in the dimer is shared by three octahedra. The only 
internal parameter $u$ controls the S-S and Co-S distances.

In Fig.~\ref{fig:CoS2_structure} we have serially numbered the four Co atoms present 
in the parent compound.  We obtained the five different Co$_{1-x}$Fe$_x$S$_2$ 
compositions for $x = 0, 0.25, 0.5, 0.75,$ and $1$ considered in the present study 
by successive replacements of Co atoms. For example, 
Co$_{0.75}$Fe$_{0.25}$S$_2$ $(x = 0.25)$ was obtained by substituting
Fe at the position Co1, Co$_{0.5}$Fe$_{0.5}$S$_2$ $(x = 0.5)$ additionally substituting
Fe at the position Co2, and so on.   
We used experimental lattice constant for the end 
member compounds CoS$_2$ and FeS$_2$~\cite{Hebert2013,Umemoto2006,Crystal}.
Experiments on intermediate compositions Co$_{1-x}$Fe$_x$S$_2$ find that the lattice 
constant follows the Vegard's law, which we utilize in our study~\cite{Wang2005,Cheng2003}.  
The internal atomic positions were relaxed for all compositions.  For the end member
compounds, the Co/Fe-S and S-S distances are parameterized by the internal parameter 
$u$, which we find to be $0.3858$ and $0.3821$ for CoS$_2$ and FeS$_2$, respectively.
There are multiple bond distances due to symmetry reduction in the intermediate 
compounds.  The experimental lattice constants and relaxed Co/Fe-S and S-S bond distances 
over the investigated range of $x$ are given in Table~\ref{Lattice parameters table}.

\section{Results and Discussion}

\begin{figure*}[!ht]
%     \centering
%     \captionsetup{width=\textwidth}    
    \includegraphics[width=0.9\textwidth,height=1.3\textwidth]{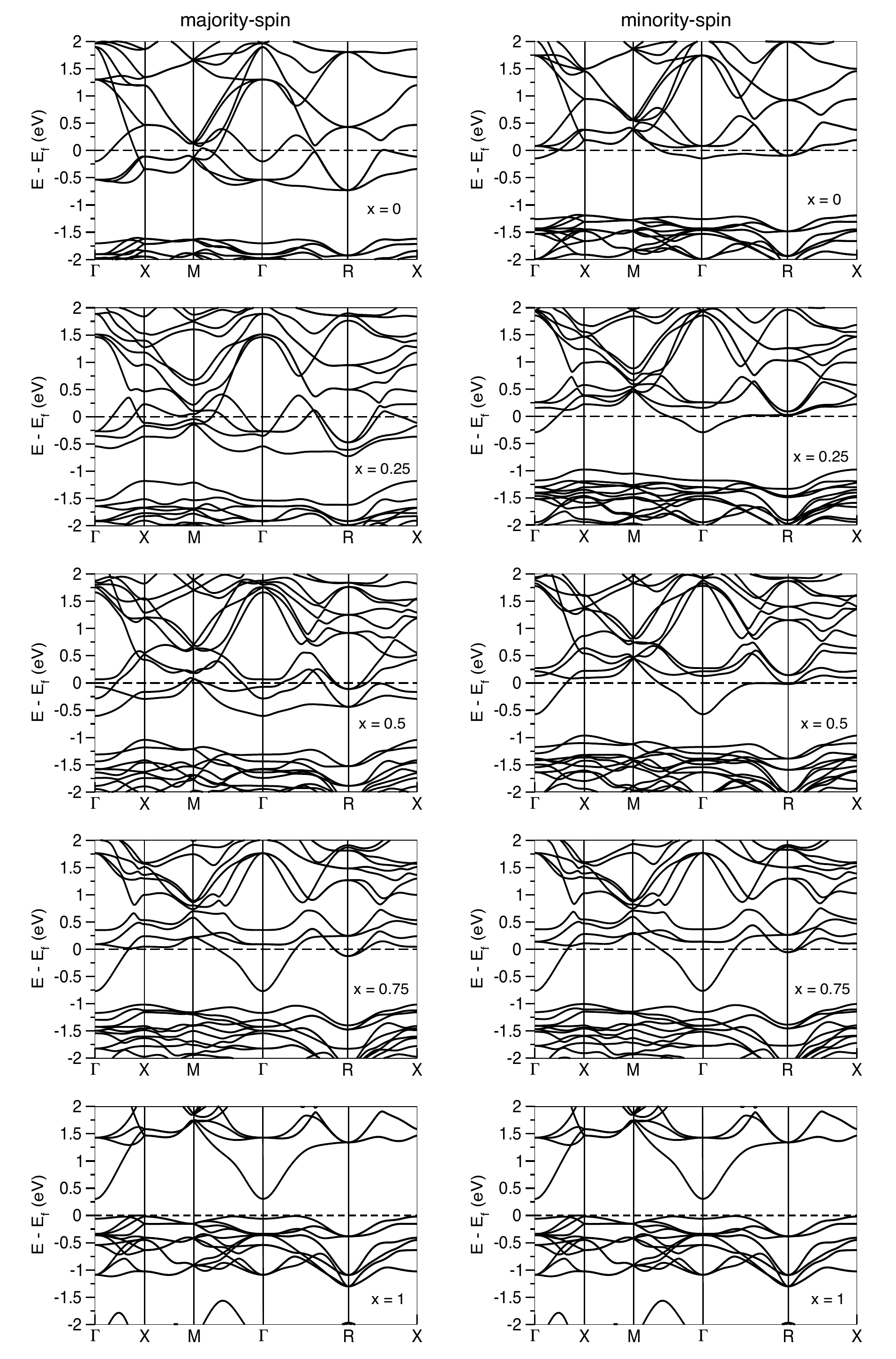}
    \caption{LDA band structures of Co$_{1-x}$Fe$_x$S$_2$ for $x = 0, 0.25, 0.5, 0.75,$ and 1. The left and right columns show the majority- and minority-spin channels, respectively. The band structure is not spin polarized for FeS$_2$ ($x = 1$).}
    \label{fig:LDA_bandstructure}
\end{figure*}

\begin{figure*}[!htbp]
%     \centering
%     \captionsetup{width=\textwidth}
    \includegraphics[width=0.9\textwidth]{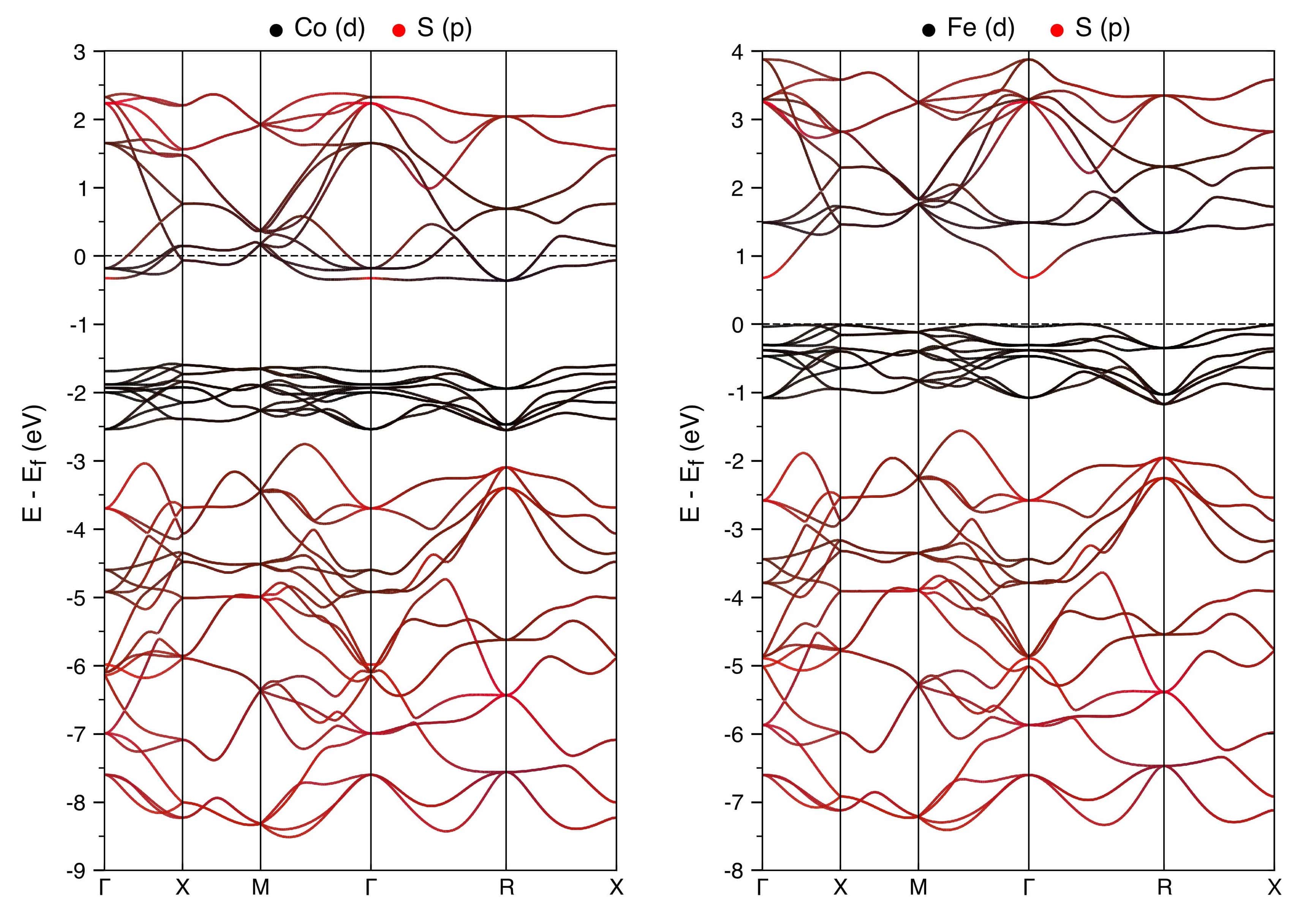}
    \caption{Non-spin-polarized band structures of CoS$_2$ (left) and FeS$_2$ (right). 
    The contribution of Co/Fe $3d$ and S $3p$ characters are depicted by black and red 
    colors, respectively. Both band structures were obtained using the experimental 
    crystal structure of FeS$_2$ to isolate the effect of chemical substitution.}
    \label{fig:Projected_band_structure}
\end{figure*}

The electronic structure of CoS$_2$ has previously been studied in detail using DFT 
calculations~\cite{Zhao1993,Shishidou2001,Umemoto2006,Feng2018,Schroter2020,Fujiwara2022}. % Bullett_1982,Kwon2000,
We briefly recapitulate the general features of the non-spin-polarized LDA 
band structure in the energy range $-9$ to 3 eV  % (not shown) 
where the S $3p$ and Co $3d$ states responsible for bonding are present. 
There is a manifold of 20 bands with predominantly S $3p$ character in the energy 
range $-9$ to $-2.5$ eV. Since there are eight S atoms in the unit cell, this 
implies that two $3p$ orbitals per S are fully filled and one $3p$ orbital per 
S takes part in the formation of S-S dimers by making a covalent bond.  Indeed, 
the four antibonding S $3p$ bands can be found above the Fermi level or partially 
crossing it.  
These 20 bands  are occupied by a total of 40 electrons, which accounts for the 32 $3p$
electrons of elemental S plus an additional 8 electrons obtained from Co $4s$ orbitals
that lie above the Fermi level. This implies nominal valences of Co$^{2+}$ and S$_2^{2-}$. 
% , since the eight electrons from
% four Co atoms are distributed over four pairs of dimerized S atoms. 
%
Just above the S $3p$ states, a manifold of 12 bands with Co $t_{2g}$ character is 
present in the range $-2.5$ to $-1.1$ eV. 
The Co $t_{2g}$ states are separated from the Co $e_g$ states by a crystal field 
splitting of around 0.4 eV.  The eight Co $e_g$ bands lie in a manifold between 
$-0.7$ and 2.5 eV, which is also shared by four antibonding S $3p$ bands.  Within 
this manifold, the low-lying bands carry predominantly Co $e_g$ character while 
the high-lying ones exhibit mostly S $3p$ character.  However, there is intermixing
of these states throughout the manifold.  In particular, the lowest-lying band
in this manifold has significant S $3p$ character near $\Gamma$. We also note 
that S $3p$ bands just below the Co $t_{2g}$ states also show some Co $3d$ character.  
This indicates significant Co $3d$-S $3p$ covalency.

We now discuss the evolution of the electronic structure near the Fermi level in
the ferromagnetic phase of these compounds as a function of Fe concentration, 
which is shown in Fig.~\ref{fig:LDA_bandstructure} for the case of LDA.  
We find that the $x = 0$ compound CoS$_2$ is not a half metal within LDA,
in agreement with previous theoretical \cite{Zhao1993, Shishidou2001}
and spin-resolved ARPES \cite{Fujiwara2022} study.  Our calculations 
using relaxed atomic positions find minority-spin electron pockets around 
both the $\Gamma$ $(0,0,0)$ and $R$ $(\frac{1}{2},\frac{1}{2},\frac{1}{2})$ 
points.  This differs from early LDA results that only find the pocket 
around $R$ \cite{Zhao1993, Shishidou2001}, but agrees with the recent
ARPES results~\cite{Wu2007,Fujiwara2022}.

\begin{figure*}[!htbp]
%     \centering
%     \captionsetup{width=\textwidth}
    \includegraphics[width=0.9\textwidth,height=1.3\textwidth]
    {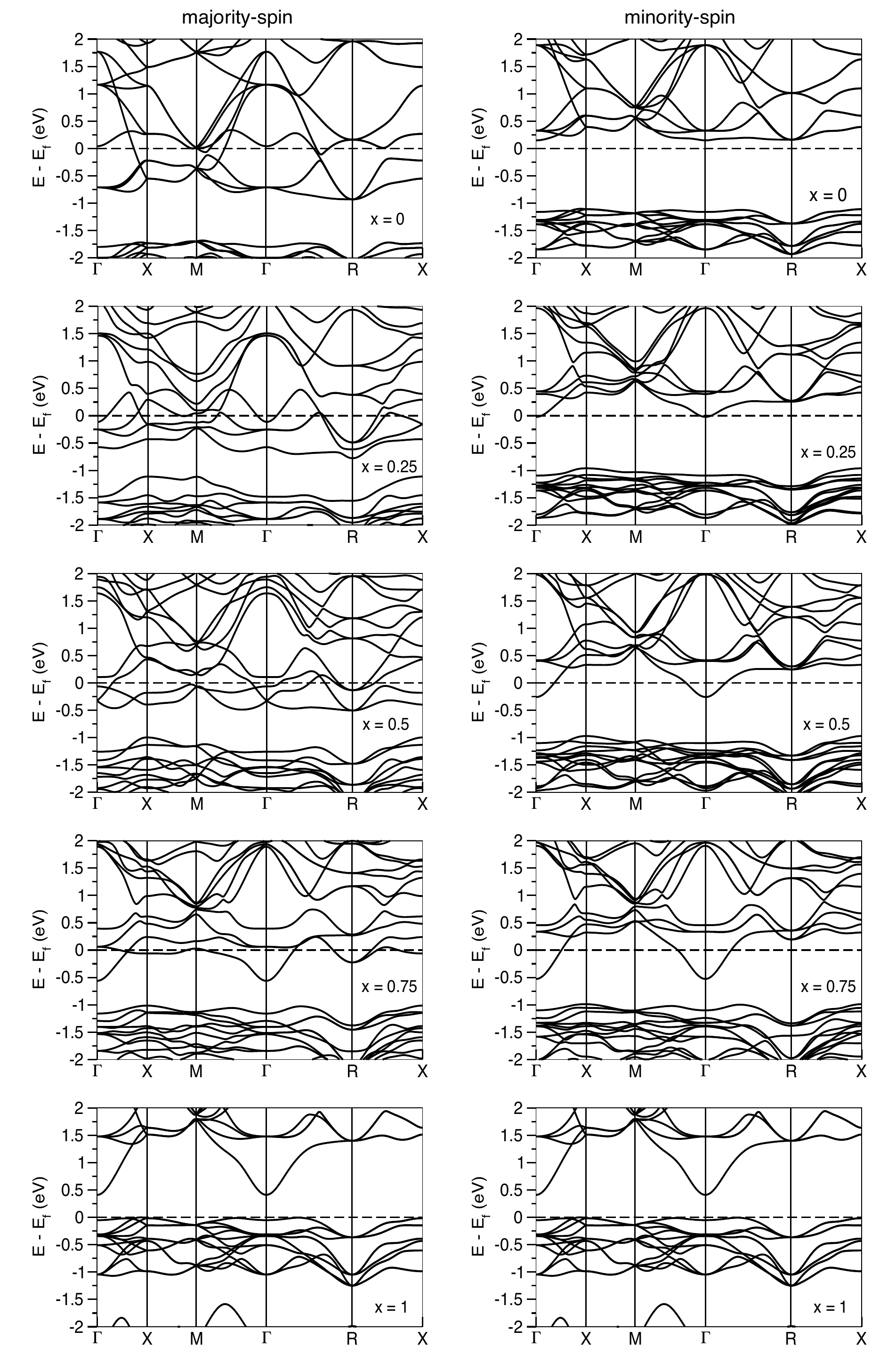}
    \caption{GGA band structures of Co$_{1-x}$Fe$_x$S$_2$ for 
    $x = 0, 0.25, 0.5, 0.75,$ and 1. The left and right columns show the 
    majority- and minority-spin channels, respectively. The band structure 
    of non-spin-polarized FeS$_2$ ($x = 1$) is shown for comparison.}
    \label{fig:PBE_bandstructure}
\end{figure*}

The band structure of Co$_{0.75}$Fe$_{0.25}$S$_2$ exhibits band splittings 
expected from lowering of symmetry due to Fe substitution.  However, the 
structure of bands that cross the Fermi level in the two spin channels
changes more than that expected from a simple downward shift of the Fermi 
level due to hole doping. In particular, the lowest-lying minority-spin 
conduction band becomes more dispersive around $\Gamma$ and dips further 
below the Fermi level.  This band has mostly S $3p$ character, and its 
dispersion increases as Fe concentration is increased.  The exchange 
splitting also decreases as more Co ions are replaced by Fe.  As a result 
the lowest-lying minority-spin band always crosses the Fermi level until 
all bands in the conduction manifold consisting of Co $e_g$ and S $3p$ 
antibonding states become unoccupied for FeS$_2$.

To understand the reason behind the increased dispersion of the lowest-lying
minority-spin conduction band as a function of Fe concentration, we examined
the band structures of CoS$_2$ and FeS$_2$ calculated using the experimental
lattice constant and atomic positions of the latter compound.  The result is 
shown in Fig.~\ref{fig:Projected_band_structure}.  We find
that the band widths of the occupied S $3p$ and Co $t_{2g}$ manifolds are 
comparable in the two compounds, with values of $\sim$6 and $\sim$1 eV, 
respectively.  However, the band width of the mixed
Co $e_g$ and S $3p$ antibonding manifold is 2.7 eV in CoS$_2$ and 3.2 eV
in FeS$_2$.  The increase in the band width is mostly due to the larger 
dispersion of the lowest- and highest-lying bands in the conduction 
manifold, which have mostly S $3p$ antibonding character.  Since this 
dispersion describes the overlap of the antibonding orbitals residing on 
neighboring S$_2^{2-}$ dimers, the reduced band width in CoS$_2$
indicates that hopping through the dimers is reduced due to screening 
of the orbitals when 
the occupation of S $3p$ antibonding states is increased. 
The reduced covalency due to increased band filling is also evident in the 
smaller distances from the midpoint of the S $3p$ manifold to the midpoints 
of the Co $t_{2g}$ and mixed Co $e_g$/S $3p$ antibonding manifolds in CoS$_2$.

Band structures of the five compounds calculated using the GGA functional, 
which are shown in  Fig.~\ref{fig:PBE_bandstructure}, also display similar 
increase of the width of the S $3p$ antibonding bands that leads 
to the occupancy of minority-spin states as Fe concentration is increased. 
CoS$_2$ is a half-metal within the GGA, consistent with previous studies \cite{Shishidou2001}.
This reflects GGA functional's tendency to overestimate magnetism.  
As electrons are removed due to Fe substitution in Co$_{0.75}$Fe$_{0.25}$S$_2$,
the lowest-lying conduction band in the minority-spin channel becomes broader
and gets slightly occupied.  The band continues to become broader as 
Fe concentration is increased, and the exchange splitting additionally
gets smaller.  As a result Co$_{1-x}$Fe$_x$S$_2$ is not half-metallic
for $x \geq 0.25$.

Table \ref{Magnetic moments table} gives total magnetization $M_s$/Co
obtained using both LDA and PBE for the five compounds that we have
considered, and the trend in $M_s$/Co highlights the role of both broadening
of the lowest-lying S $3p$ antibonding band and reduction of the exchange splitting 
in forestalling the formation of a half-metallic state as Fe concentration 
is increased.  In agreement with the experiments, $M_s$/Co obtained using 
LDA increases going from $x = 0$ to 0.25,  indicating that the electronic 
states are more exchange-split in Co$_{0.75}$Fe$_{0.25}$S$_2$.  This is also 
evident when one compares the LDA band structures for $x = 0$ and 0.25 in 
Fig.~\ref{fig:LDA_bandstructure}. The flat part of the lowest-lying minority-spin 
 conduction band along $\Gamma$-$R$ is below the Fermi 
level for $x = 0$.  However, when this band becomes more dispersive around
$\Gamma$ for $x = 0.25$, the flat part along $\Gamma$-$R$ shifts slightly
above the Fermi level in the minority-spin channel, which results in an 
increased value of $M_s$/Co. Similarly, $M_s$/Co obtained from PBE decreases 
by only 0.2\% when going from $x = 0$ to 0.25.  The corresponding PBE band 
structures in Fig.~\ref{fig:PBE_bandstructure} again show that the main effect
of Fe substitution is to increase the dispersion of the lowest-lying conduction 
band in the minority channel around $\Gamma$, which consists of states with 
predominantly S $3p$ antibonding character.  The broader S $3p$ antibonding 
band slightly crosses the Fermi level, once more highlighting that 
occupation of minority-spin states for $x = 0.25$ occurs due to changes in 
screening of the S $3p$ antibonding orbitals rather than due to a decrease in 
exchange splitting. As Fe concentration is further increased, the relative 
splitting of the conduction manifold in the majority- and minority-spin 
channels also decreases significantly within both LDA and GGA.  This leads to 
increasing occupation of the minority-spin states, and it gets reflected in 
the noticeably smaller $M_s$/Co obtained using both LDA and GGA as $x$ is increased 
to 0.5 and 0.75.   We note that the nonintegral value of $M_s$/Co we obtain is 
in contrast to previous calculations that find $M_s$/Co = 1$\mu_B$ \cite{Day-Roberts2020,Wang2005,Ramesha2004,Cheng2003},
which is 
likely caused by the use of experimental internal atomic positions, the DFT+$U$ method, or fully relaxed lattice parameters. However,  experiments \cite{Jarrett1968, Ogawa1974}
do observe a decrease in $M_s$/Co value for higher Fe concentrations, which is also supported by theoretical results \cite{Mazin2000}.

\begin{table}[!htbp]
% \captionsetup{width=0.9\textwidth}
\caption{Magnetic moments per Co of Co$_{1-x}$Fe$_x$S$_2$ obtained using LDA and GGA.
}
\label{Magnetic moments table}
\begin{ruledtabular}
\begin{tabular}{l c c c c}
Compound & LDA & PBE \\ 
    &  $M_{s}$/Co ($\mu_{B}$) & $M_{s}$/Co ($\mu_{B}$)\\ \hline
CoS$_2$ & 0.720 & 1.0 \\
Co$_{0.75}$Fe$_{0.25}$S$_2$ & 0.915 & 0.998\\
Co$_{0.5}$Fe$_{0.5}$S$_2$ & 0.715 & 0.965 \\
Co$_{0.25}$Fe$_{0.75}$S$_2$ & 0.179 & 0.803 \\
FeS$_2$ & $-$ & $-$ \\
\end{tabular}
\end{ruledtabular}
\end{table}

\section{Conclusions}

We investigated the electronic and magnetic properties of Co$_{1-x}$Fe$_x$S$_2$
for $x = 0, 0.25, 0.5, 0.75$, and 1 using first principles calculations 
and showed that this series of compounds does not host a half-metallic
phase.  The effect of chemical doping was considered by explicit 
substitution of Co with Fe.  We find that the details of the band structure 
of these compounds obtained using LDA and GGA functionals are different.  
However, both functionals find that the lowest-lying minority-spin conduction
band becomes more dispersive as Fe concentration is increased such that this 
band remains partially occupied even when the overall band filling is decreased 
via hole doping. Our band structure calculations for CoS$_2$ and FeS$_2$ using 
the same structural parameters highlight that the increase in band width of the 
conduction manifold is due to reduced screening of the S $3p$ antibonding states.  
Furthermore, we find that the exchange splitting between the majority- and 
minority-spin states decrease for compounds with higher Fe concentrations.  
As a result, the Fermi level and the bands in the minority-spin channel 
simultaneously get lowered, and there is a finite occupation of minority-spin 
states at the Fermi level for all the nonzero values of Fe concentrations that 
we studied.  Since ARPES experiments show that the lowest-lying minority-spin 
conduction band crosses the Fermi level even in stoichiometric CoS$_2$, the 
two trends identified in our results indicate that Co$_{1-x}$Fe$_x$S$_2$ may 
not be half metallic for any Fe concentration. 

\begin{acknowledgements}
We are thankful to Sylvie H\'ebert for useful discussions.  This work was 
supported by Agence Nationale de la Recherche under grant ANR-21-CE50-0033
and GENCI-TGCC under grant A0130913028. 
\end{acknowledgements}

\bibliography{references}

\end{document}